\documentclass[aps,draft,12pt]{revtex4-1}

\usepackage{graphicx}
\usepackage{dcolumn}
\usepackage{bm}

\begin{document}

\preprint{}

\title{A mechanism for formation of Bose-Einstein condensation in cosmology}

\author{Recai Erdem}
\email{recaierdem@iyte.edu.tr}
\author{Kemal G{\"{u}}ltekin}
\email{kemalgultekin@iyte.edu.tr}
\affiliation{Department of Physics \\
{\.{I}}zmir Institute of Technology\\
G{\"{u}}lbah{\c{c}}e, Urla 35430, {\.{I}}zmir, Turkey}

\date{\today}

\begin{abstract}
We introduce a toy model of scalar particles with a trilinear scalar coupling in cosmology. The trilinear coupling $\phi^2\chi$ causes production of  non-relativistic $\phi$ particles through the process $\chi\chi\,\rightarrow\,\phi\phi$ where, initially, only relativistic $\chi$ particles are present. We consider the initial times of $\chi\chi\,\rightarrow\,\phi\phi$ and observe that the curved space effects promote formation of Bose-Einstein condensate of $\phi$ particles.
\end{abstract}



\maketitle

\section{Introduction}

Observations suggest that  $\sim\,95\,\%$ of the universe consists of non-luminous energy densities, namely, dark energy and dark matter, and $\sim\,5\%$ consists of the usual matter (i.e. baryonic matter and radiation), and initially we had a era of cosmological exponential expansion, namely, cosmological inflation. The current standard model that describes the observational picture at present is $\Lambda$CDM where dark energy is described by cosmological constant and dark matter is described by cold (i.e. low kinetic energy) dust that is made of dark particles, respectively \cite{Weinberg}.  $\Lambda$CDM has some serious problems. For example, if cosmological constant is adopted as dark energy then the cosmological constant problem arises \cite{ccp}.  It seems that the explanation of dark matter by cold ordinary type of particles also have some problems, such as, the problem of predicting too dense cores for the galaxies (i.e. core-cusp or cuspy halo problem), too many dwarf galaxies when compared with observations \cite{CDM}. One of the most popular and preferred alternatives to $\Lambda$CDM are those that employ scalar fields for dark energy and dark matter. The standard framework for cosmic inflation is already a scalar field (or scalar fields) called inflaton. In this type of models the scalar depends only on time, and only its perturbations (that correspond to cosmological perturbations) depends on the usual 3-dimensional coordinates (and time). This is a natural situation if the scalar field corresponds to a Bose-Einstein condensate state. This view is also supported by the equivalence of $\phi^4$ theories at relativistic level to the Gross-Pitaevskii equation of Bose-Einstein condensation at non-relativistic level \cite{Morikawa-2,BEC-phi4,Erdem}. Many models of Bose-Einstein condensate scalar fields for dark matter, dark energy, and inflation are studied in literature \cite{BEC-DM,BEC-DE,Morikawa-2, BEC-inflation}. These models are mainly interested in manifestations of the Bose-Einstein condensation while the conditions for formation of the condensate are studied roughly. In fact the formation of the Bose-Einstein condensation (BEC) is not wholly understood yet in the sense that in all BEC formation studies one starts with an initial, already formed seed of BEC \cite{BEC-formation,kinetic-theory}. Moreover, in all these models the formation and the evolution of the condensation is mainly studied at macroscopic level (i.e. at the level of number densities or distribution functions) while the dynamics of the condensation at microscopic level (i.e. at the level of particle physics) is not considered sufficiently.

The aim of this paper is to study the initial phase of formation of a scalar Bose-Einstein condensation in cosmology with particular emphasis on its microscopic description in particle physics. In our model formation and evolution of the condensate are induced by a $\phi^2\chi$ term in the Lagrangian. There are other studies in literature that study the effect of curved space on the decay rates for scalar fields conformally coupled to gravity \cite{mass-generation} while, in this paper, we study the situation for scalars minimally coupled to gravity in an approximate cosmological setting that satisfy some relevant conditions. The rest of this paper is organized as follows. In Section II we give the essential elements of the model. In Section III we show that all necessary pre-conditions for the formation of the condensation are satisfied in this model. In Section IV we conclude, and the technical details are studied in the appendices.

\section{The Model and the Scheme}

We take the Robertson-Walker metric
\begin{equation}
ds^2\,=\,-dt^2\,+\,a^2(t)[\frac{dr^2}{1-kr^2}+r^2
(d\theta^2+\sin^2{\theta}d\phi^2)] \label{e1}
\end{equation}
and for simplicity we let $k=0$ which is in agreement with
observations \cite{PDG}.
We consider the following action in this space
\begin{eqnarray}
&S&
\,=\,\int\sqrt{-g}\;d^4x\,\frac{1}{2}\{-g^{\mu\nu}\left[\partial_\mu\phi\partial_\nu\phi\,+\,
\partial_\mu\chi\partial_\nu\chi\right]\,-\,m_\phi^2\phi^2\,-\,
m_\chi^2\chi^2\,-\,\mu\,\phi^2\chi\} \label{e2a} \\
&=&
\int\,d^3x\,d\eta\,\frac{1}{2}\{\,
\tilde{\phi}^{\prime\;2}-(\vec{\nabla}\tilde{\phi})^2
+\tilde{\chi}^{\prime\;2}-(\vec{\nabla}\tilde{\chi})^2
-\tilde{m}_\phi^2\tilde{\phi}^2-\,
\tilde{m}_\chi^2\tilde{\chi}^2
-\tilde{\mu}\,\tilde{\phi}^2\tilde{\chi}\,\} \;.
\label{e2b}
\end{eqnarray}
Here prime denotes derivative with respect to conformal time $\eta$ \cite{QFTC} and
\begin{eqnarray}
&&d\eta\,=\,\frac{dt}{a(t)}~,
~\tilde{\phi}\,=\,a\phi~,
~\tilde{\chi}\,=\,a\chi~,~a^\prime\,=\,\frac{da}{d\eta}~,~\dot{a}\,=\,\frac{da}{dt}~,~a^{\prime\prime}\,=\,\frac{d^2a}{d\eta^2}
~,~\ddot{a}\,=\,\frac{d^2a}{dt^2}, \nonumber \\
&&\tilde{\mu}\,=\,a\,\mu~,~~\tilde{m}_i^2\,=\,m_i^2a^2-\frac{a^{\prime\prime}}{a}\,=\,a^2\left(m_i^2-\frac{\ddot{a}}{a}-\frac{\dot{a}^2}{a^2}\right)\;,
\label{e2aa}
\end{eqnarray}
where the subscript $i$ takes the values, $i\,=\,\phi,\chi$, and a dot over a quantity denotes its derivative with respect to time $t$.

\begin{figure}[h]
\begin{picture}(200,150)(50,50)
\put (0,60){\vector(1,-1){15}}
\put (15,45){\line(1,-1){15}}
\put (30,30){\vector(1,1){15}}
\put (45,45){\line(1,1){15}}
\put (0,-60){\vector(1,1){15}}
\put (15,-45){\line(1,1){15}}
\put (30,-30){\vector(1,-1){15}}
\put (45,-45){\line(1,-1){15}}
\put(5,40){$\tilde{p}_1$}
\put(20,45){$\chi$}
\put(5,-40){$\tilde{p}_2$}
\put(20,-45){$\chi$}
\put(60,50){$\tilde{p}_3$}
\put(40,50){$\phi$}
\put(60,-50){$\tilde{p}_4$}
\put(40,-50){$\phi$}
{\bf {\linethickness{1pt}\put(30,-30){\line(0,1){60}}}}
\put(20,5){$\tilde{q}_t$}
\put(35,0){$\phi$}
\put(35,-80){(a)}
\hspace*{200pt}
\put (0,60){\vector(1,-1){15}}
\put (15,45){\line(1,-1){15}}
\put (30,30){\vector(1,-1){15}}
\put (45,15){\line(1,-1){40}}
\put (0,-60){\vector(1,1){15}}
\put (15,-45){\line(1,1){15}}
\put (30,-30){\vector(1,1){15}}
\put (45,-15){\line(1,1){40}}
\put(5,40){$\tilde{p}_1$}
\put(20,45){$\chi$}
\put(5,-40){$\tilde{p}_2$}
\put(20,-45){$\chi$}
\put(65,18){$\tilde{p}_3$}
\put(50,25){$\phi$}
\put(65,-19){$\tilde{p}_4$}
\put(50,-30){$\phi$}
{\bf {\linethickness{1pt}\put(30,-30){\line(0,1){60}}}}
\put(20,5){$\tilde{q}_u$}
\put(35,0){$\phi$}
\put(35,-80){(b)}
\end{picture}
\hspace{10pt}\\
\hspace{10pt}\\
\hspace{10pt}\\
\hspace{10pt}\\
\hspace{10pt}\\
\hspace{10pt}\\
\hspace{10pt}\\
\caption{The leading order Feynman diagrams that may contribute to the production of $\phi$ particles. Here
$\tilde{q}_t\,=\,\tilde{p}_1-\tilde{p}_3=\tilde{p}_4-\tilde{p}_2$,
$\tilde{q}_u\,=\,\tilde{p}_1-\tilde{p}_4\,=\tilde{p}_3-\tilde{p}_2$ are the 4-momenta carried in the internal lines. Note that the s-channel is forbidden in this case by kinematics.}
\label{fig:1}
\end{figure}
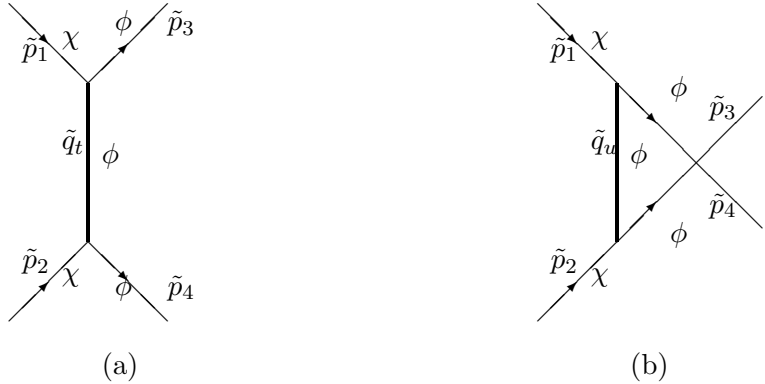

We take $\frac{\tilde{\mu}}{\tilde{m}_\phi}\,\ll\,1$, $\tilde{m}_\chi\,\ll\,\tilde{m}_\phi$, and assume that initially there are only $\chi$ particles (that may be identified by inflaton, curvaton or a decay product of inflaton). Then the leading order contributions to the production of $\phi$ particles are given in Figure \ref{fig:1}. Furthermore we assume that the rate of $\chi\chi\,\rightarrow\,\phi\phi$ is much larger than the Hubble parameter at the time of this process so that one may take the masses $\tilde{m}_\chi$, $\tilde{m}_\phi$ constant during each of the processes $\chi\chi\,\rightarrow\,\phi\phi$ while the time dependence is observed only at cosmological scales.
This condition may be expressed as the variation in $\tilde{m}_\chi^2=a^2\left(m_\chi^2-\dot{H}-2H^2\right)$ and $\tilde{m}_\phi^2=a^2\left(m_\phi^2-\dot{H}-2H^2\right)$  for a single process for $\chi\chi\,\rightarrow\,\phi\phi$ during the time $\Delta\,t=\frac{1}{n_\chi\beta\sigma\,v}$ should be very small i.e.
\begin{equation}
\frac{\Delta\,\tilde{m}^2}{\tilde{m}^2}\,=\,\left|\frac{\left(\frac{1}{n_\chi\beta\sigma\,v}\right)\left(\frac{d\,a^2\left(m^2-\dot{H}-2H^2\right)}{dt}\right)}{a^2\left(m^2-\dot{H}-2H^2\right)}\right|\,\ll\,1 \;,\label{t1x}
\end{equation}
where $m$ denotes either of $m_\chi$ or $m_\phi$, $\dot{H}=\frac{dH}{dt}=\frac{\ddot{a}}{a}-H^2$, $\beta$ is the effective penetration depth of the incoming beam to the target, $n_\chi$ is the number density of the target particles, $\sigma$ is the cross section of the process, $v$ is the relative velocity of the incoming and the target particles.
In fact, one may impose $\frac{\Delta\,\tilde{m}}{\tilde{m}}\,\ll\,1$ in a smaller time interval $\Delta\,t\,\leq\,\frac{1}{n_\chi\beta\sigma\,v}$ whenever (\ref{t1x}) is not satisfied. However, in the case $\Delta\,t\,\leq\,\frac{1}{n_\chi\beta\sigma\,v}$, one may use this formulation only for the modes with sufficiently shorter wavelengths. But, in any case, one can not use this formulation for $\Delta\,t\,\geq\,\frac{1}{n_\chi\beta\sigma\,v}$. In that case, the fields can not be taken to be asymptotically free, and
the modes with larger wavelengths than the separation of two particles become correlated in any case (and Bose-Einstein condensation may develop), so this formulation would become inapplicable since all modes would behave as a single entity and the present formulation would not apply. Therefore we adopt $\Delta\,t=\frac{1}{n_\chi\beta\sigma\,v}$ rather than a smaller or larger $\Delta\,t$. In fact, as is shown in Appendix A, there is a considerably large, phenomenologically relevant parameter space where (\ref{t1x}) is satisfied.

The quantum field theory in Minkowski space for interacting particles relies on the expression of interacting quantum fields in terms of interaction picture fields that evolve as free fields (with constant masses) \cite{QFT}. This condition is satisfied in each time interval provided that (\ref{t1x}). The standard formulation of quantum field theory in Minkowski space also requires constancy of $\tilde{\mu}$ in each interval $\eta_i\,<\,\eta\,<\,\eta_{i+1}$, and this may be easily imposed by requiring
\begin{equation}
\left|\frac{\Delta\,\tilde{\mu}}{\tilde{\mu}}\right|\,=\,\left|\frac{\frac{d\tilde{\mu}}{dt}\Delta\,t}{\tilde{\mu}}\right|
\,=\,\left|H\,\Delta\,t\right|\,=\,\left|\frac{H}{n_\chi\beta\sigma\,v}\right|\,\ll\,1 \;.\label{r2a}
\end{equation}
One may show that, for a space where the Hubble parameter and the scale factor are related by some function $f$ i.e. for $H=f(a)$, the condition (\ref{r2a}) guarantees the condition (\ref{t1x}), provided that $\frac{df}{da}$ and $\frac{d^2f}{da^2}$ are not extremely large (and if $\left|\frac{H}{n_\chi\beta\sigma\,v}\right|$ is sufficiently small) while (\ref{t1x}) does not guarrantee (\ref{r2a}).  For example, in the case of the phenomenologically relevant simple cases in Appendix A, the condition (\ref{r2a}) guarantees $\left|\frac{2H}{n_\chi\beta\sigma\,v}\right|\left|1-\frac{s(s-2)H^2}{m_\chi^2+(s-2)H^2}\right|\ll\,1$ i.e. the condition corresponding to (\ref{t1x}), namely, (\ref{t1xxx}) for reasonable values of parameters while (\ref{t1xxx}) does not guarantee  (\ref{r2a}). On the other hand, the condition (\ref{t1xxx}) has a wider scope since it may also be guaranteed by imposing $\left|\left(1-\frac{s(s-2)H^2}{m_\chi^2+(s-2)H^2}\right)\right|\,\ll\,1$ while (\ref{r2a}) is much simpler. However, the additional possibility of $\left|\left(1-\frac{s(s-2)H^2}{m_\chi^2+(s-2)H^2}\right)\right|\,\ll\,1$ does not expand the physically allowed region considerably as shown in Appendix A. In other words, the implications of (\ref{t1x}) and (\ref{r2a}) are similar while they are not wholly equivalent.

 The condition (\ref{t1x}) insures that the corresponding masses may be taken to be constant, and (\ref{r2a}) insures that the coupling constant may taken to be constant in each time interval $\Delta\,t$.
 Therefore, one may use the tools of the usual perturbative quantum field theory for calculation of the rates and cross sections in an effective Minkowski space e.g. for each of the process given in Figure \ref{fig:1} if the conditions (\ref{t1x}) and (\ref{r2a}) are satisfied. One may take the masses and the coupling constant be constant during a process and use the usual formulas for the rates and cross sections of the usual (Minkowski space) quantum field theory, and then one may take masses and the coupling constant of the particles during the next process be other constants, and then calculate the rates and cross sections for that process, and so on. The conditions (\ref{t1x}) and (\ref{r2a}) insure that a possible contribution to decay widths \cite{mass-generation} and gravitational particle production \cite{GPP} due to change in the effective masses and the coupling constant is small \cite{QFTC} in each time interval.

 The main result of this section may be summarized as follows: One may consider the metric (\ref{e1}) Minkowskian in each time interval $\Delta\,t$ provided that (\ref{t1x}) and (\ref{r2a}) are satisfied. Therefore any field $\chi$ that satisfies (\ref{t1x}) and (\ref{r2a}) may be expressed in the $i$th time interval as (see Appendix B for a rigorous derivation)
\begin{eqnarray}
\tilde{\chi}^{(i)}(\vec{r},\eta)\,\simeq\,
\int\,
\frac{d^3\tilde{p}}{(2\pi)^\frac{3}{2}\sqrt{2\omega_p^{(i)}}}\left[a_p^{(i)\,-}\,
e^{i\left(\vec{\tilde{p}}.\vec{r}-\omega_p^{(i)}(\eta-\eta_i)\right)}
\,+\,a_p^{(i)\,+}
\,e^{i\left(-\vec{\tilde{p}}.\vec{r}+\omega_p^{(i)}(\eta-\eta_i)\right)}\right] \;,\label{e2aaaxx}
\end{eqnarray}
where $\eta_i\,<\,\eta\,<\,\eta_{i+1}$, and the superscript $^{(i)}$ refers to the $i$th time interval between the $i$th and $(i+1)$th processes.
In other words, Eqs. (\ref{t1x}) and (\ref{r2a}), and the form of (\ref{e2b}) imply that we have an effective Minkowski space given by \cite{Parker}
\begin{equation}
d\tilde{s}^2\,=\,-d\eta^2\,+\,d\tilde{x}_1^2+d\tilde{x}_2^2+d\tilde{x}_3^2 \label{t1ax}
\end{equation}
in each interval $\eta_i\,<\,\eta\,<\,\eta_{i+1}$, for $\chi$, where the masses and the coupling constant of the particles are constant and they alter as one passes from one interval to the other, and $\tilde{x}_i$ are related to (\ref{e1}) by $d\tilde{x}_1^2+d\tilde{x}_2^2+d\tilde{x}_3^2$=$dr^2+r^2(d\theta^2+\sin^2{\theta}d\phi^2)$. Therefore, one may employ the tools of the usual perturbative quantum field theory in each interval in (\ref{t1ax}), as will be done in the next section.

\section{Realizing the conditions for condensation}
\subsection{Achieving coherence and correlation}

The phase space evolution of $\phi$ particles may be determined by the evolution of the distribution function for one of the final state particles \cite{kinetic-theory} through the equation
\begin{eqnarray}
\frac{d\,\tilde{f}(\vec{\tilde{p}}_4,\eta)}{d\eta}&=&\frac{1}{32(2\pi)^5\tilde{E}_4}\int\int\int\,\delta^{(4)}(\tilde{p}_1+\tilde{p}_2-\tilde{p}_3-\tilde{p}_4)
\,|\tilde{M}|^2\nonumber \\
&&\times\,\{\tilde{f}_1\tilde{f}_2(1+\tilde{f}_3)(1+\tilde{f}_4)-\tilde{f}_3\tilde{f}_4(1+\tilde{f}_1)(1+\tilde{f}_2)\}
\frac{d^3\vec{\tilde{p}}_1}{\tilde{E}_1}\frac{d^3\vec{\tilde{p}}_2}{\tilde{E}_2}\frac{d^3\vec{\tilde{p}}_3}{\tilde{E}_3}\;, \label{t1}
\end{eqnarray}
where the notation $d^3\vec{p}$ is used (rather than $d^3p$) to prevent any possible confusion between 3-vector $\vec{p}$ and 4-vector $p$ while we prefer the notation $d^3p$ when there is no danger of confusion, and  $\tilde{}$ refers to the effective Minkowski space (\ref{t1ax}), $\tilde{f}_i=\tilde{f}(\vec{\tilde{p}}_i,\eta)$ is the number density in phase space corresponding to (\ref{t1ax}). Here $|\vec{\tilde{p}}|$ is
\begin{equation}
|\vec{\tilde{p}}|\,=\,\tilde{m}_\chi\sqrt{\tilde{g}_{ij}\frac{d\tilde{x}^i}{d\tilde{\tau}}\frac{d\tilde{x}^j}{d\tilde{\tau}}}\,=\,
a\,m_\chi\sqrt{g_{ij}\frac{dx^i}{d\tau}\frac{dx^j}{d\tau}}\,=\,a |\vec{p}| \;. \label{t1y}
\end{equation}
In the above, tilde $\tilde{}$ over a quantity refers to its form for the metric (\ref{t1ax}) while the quantities without a tilde refer to its form for the metric (\ref{e1}), and $d\tau^2=-ds^2$, $d\tilde{\tau}^2=-d\tilde{s}^2$. To derive Eq. (\ref{t1y}) we use $g_{ij}= a^2\tilde{g}_{ij}$ and $\frac{d\tilde{x}^i}{d\tilde{\tau}}=a^2\left(\frac{m}{\tilde{m}}\right)\frac{dx^i}{d\tau}$ which, in turn, follows from the geodesic equations for the actions $\int\,\tilde{m}_\chi\,\sqrt{-\tilde{g}_{\mu\nu}\frac{d\tilde{x}^\mu}{d\tilde{\tau}}\frac{d\tilde{x}^\nu}{d\tilde{\tau}}}\,d\tilde{\tau}$ and
$\int\,m_\chi\,\sqrt{-g_{\mu\nu}\frac{dx^\mu}{d\tau}\frac{dx^\nu}{d\tau}}\,d\tau$ for the metrics (\ref{t1ax}) and (\ref{e1}), respectively, after requiring that $|\vec{\tilde{p}}|$ and $|\vec{p}|$ coincide for $a=\mbox{constant}=1$.
We observe that $|\vec{p}|\propto\,\frac{1}{a}$ is the physical momentum while $|\vec{\tilde{p}}|$ does not depend on redshift.

$\tilde{M}$ in (\ref{t1}) denotes the transition matrix element for $\chi\chi\,\rightarrow\,\phi\phi$  which is dominated by the tree-level diagrams in Figure \ref{fig:1}. The $\tilde{M}$ corresponding to Figure \ref{fig:1} is given by
\begin{equation}
\tilde{M}\,=\,
\tilde{\mu}^2\left[
\frac{1}{(\tilde{p}_1-\tilde{p}_3)^2+\tilde{m}_\phi^2}
\,+\,\frac{1}{(\tilde{p}_1-\tilde{p}_4)^2+\tilde{m}_\phi^2}\right]\;.
\label{t1b}
\end{equation}

The number density in phase space for (\ref{t1ax}), namely, $\tilde{f}_i=\tilde{f}(\vec{\tilde{p}}_i,\eta)$ may be related to the one for (\ref{e1}), $f_i=f(\vec{p}_i,t)$ as
\begin{equation}
\tilde{f}_i\,=\,\tilde{f}(\vec{\tilde{p}}_i,\eta)\,=\,\frac{d\tilde{N}(\eta)}{d^3\tilde{p}_{(i)}d^3\tilde{x}_{(i)}}\,=\,\frac{d\tilde{N}(\eta)}{d^3p_{(i)}\,d^3x_{(i)}}
\,=\,\frac{dN(t)}{d^3p_{(i)}\,d^3x_{(i)}}\,=\,f(\vec{p}_i,t)\,=\,f_i\;, \label{s1}
\end{equation}
 where the sub-index $(i)$ refers to the i'th particle, and we have used $\vec{p}=\frac{1}{a}\vec{\tilde{p}}$, $\vec{x}=a\vec{\tilde{x}}$ where $x$ is the physical length scale. Hence, (\ref{t1}), in terms of the quantities corresponding to (\ref{e1}), reads
\begin{eqnarray}
\frac{df(\vec{p}_4,t)}{dt}&=&\frac{a^4(t)}{32(2\pi)^5E_4}\int\int\int\,\delta^{(4)}(p_1+p_2-p_3-p_4)
\,|M|^2\nonumber \\
&&\times\,\{f_1f_2(1+f_3)(1+f_4)-f_3f_4(1+f_1)(1+f_2)\}
\frac{d^3\vec{p}_1}{E_1}\frac{d^3\vec{p}_2}{E_2}\frac{d^3\vec{p}_3}{E_3} \;.\label{t1s}
\end{eqnarray}
Here we have used (\ref{s1}), and $\frac{d}{d\eta}=a\frac{d}{dt}$, $\vec{\tilde{p}}=a\vec{p}$, $\delta(ax)=\frac{1}{a}\delta(x)$; $M$ is obtained from $\tilde{M}$ by replacing the quantities corresponding to (\ref{t1ax}) by those corresponding to (\ref{e1}) while their numerical values are the same i.e.  $M=\tilde{M}$; $E$ is found from $\tilde{E}$ by replacing the quantities corresponding to (\ref{t1ax}) by those corresponding to (\ref{e1}) and then multiplying it by $\frac{1}{a}$ i.e.
\begin{equation}
E_i^2\,=\,\left(m_i^2-\frac{\ddot{a}}{a}-\frac{\dot{a}^2}{a^2}\right)+\vec{p}_i^2~~~\mbox{while}~~~
\tilde{E}_i^2\,=\,a^2m_i^2-\frac{a^{\prime\prime}}{a}+\vec{\tilde{p}}^2\;. \label{s2}
\end{equation}
In the following paragraphs of this section we prefer to use (\ref{t1}) because of its simplicity while we will use (\ref{t1s}) in the following section.

In the center of mass frame, conservation of energy amounts to
\begin{equation}
\vec{\tilde{p}}^2+\tilde{m}_\chi^2=\vec{\tilde{k}}^2+\tilde{m}_\phi^2~~~\mbox{i.e.}~~~\vec{\tilde{p}}^2-\vec{\tilde{k}}^2=\tilde{m}_\phi^2 -\tilde{m}_\chi^2=a^2\left(m_\phi^2 -m_\chi^2\right) \;,\label{x}
\end{equation}
where $\vec{\tilde{k}}=\vec{\tilde{p}}_3$ and $\vec{\tilde{p}}=\vec{\tilde{p}}_1$ in the center-of-mass system. Note that $\vec{\tilde{p}}$, $\vec{\tilde{k}}$ do not depend on time, so, if (\ref{x}) is satisfied for a value of $\vec{\tilde{p}}^2-\vec{\tilde{k}}^2$ (at the moment of some transition $\chi\chi\,\rightarrow\,\phi\phi$) then, in general, it is not satisfied at a later time by the same $\vec{\tilde{p}}^2-\vec{\tilde{k}}^2$. For a given value of $|\vec{\tilde{p}}|$, the corresponding $|\vec{\tilde{k}}|$ will get smaller and smaller by time. In fact, this observation is one of the key points for deriving the tendency of the $\phi$s in this study towards Bose-Einstein condensation as we will see in (\ref{t2vx}).

In this study we assume that the momentum of the $\chi$ particles (that have enough energies to induce $\chi\chi\,\rightarrow\,\phi\phi$ processes) satisfies $0\,\leq\,|\vec{\tilde{p}}|_{min}\,<\,|\vec{\tilde{p}}|\,<\,|\vec{\tilde{p}}|_{max}$, and the spatial distributions of the $\chi$ particles in this range are homogeneous and isotropic. This condition together with the fact that $\tilde{n}_\chi(\eta)=\int\,d^3\tilde{p}\,\tilde{f}^{(i)}(\vec{\tilde{\tilde{p}}},\eta)$, (where $\tilde{n}_\chi$ is the number density of $\chi$ in the co-moving frame defined by (\ref{t1ax})), in turn, suggests that
\begin{eqnarray}
&&\tilde{f}^{(i)}(\vec{\tilde{p}}_j,\eta)\,\simeq\,\tilde{n}_\chi\,
\frac{\left[\Theta\left(|\vec{\tilde{p}}|_{max}-|\vec{\tilde{p}}|_{min}\right)-
\Theta\left(|\vec{\tilde{p}}|_{min}-|\vec{\tilde{p}}|_{max}\right)\right]}{4\pi\left(|\vec{\tilde{p}}|_{max}|-
|\vec{\tilde{p}}|_{min}\right)|\vec{\tilde{p}}_j|^2}
\left[\Theta\left(|\vec{\tilde{p}}_j|-|\vec{\tilde{p}}|_{min}\right)-
\Theta\left(|\vec{\tilde{p}}_j|-|\vec{\tilde{p}}|_{max}\right)\right]\;, \nonumber\\
&&\label{t3a}
\end{eqnarray}
where $\Theta$ denotes Heaviside function (i.e. unit step function), $j=1,2$. In fact, provided that, we impose  $|\vec{\tilde{p}}|_{max}\,>\,|\vec{\tilde{p}}|_{min}\,\geq\,0$, the
$\left[\Theta\left(|\vec{\tilde{p}}|_{max}-|\vec{\tilde{p}}|_{min}\right)-
\Theta\left(|\vec{\tilde{p}}|_{min}-|\vec{\tilde{p}}|_{max}\right)\right]$ term in (\ref{t3a}) may be set to $1$. However we keep that term since it makes (\ref{t3a}) an even function in an analytical way, this, in turn, makes the identification of the delta function in (\ref{t2vx}) easier as we shall see. Another important comment is in order here; although (\ref{t3a}) may not be the only choice that is isotropic and compatible with $\tilde{n}_\chi(\eta)=\int\,d^3\tilde{p}\,\tilde{f}^{(i)}(\vec{\tilde{\tilde{p}}},\eta)$,
$\tilde{f}^{(i)}$ given in (\ref{t3a}) has the virtue of being in the form of distribution function of the density perturbations of inflaton (or other scalar fields) in the inflationary era. In inflationary models the form of the classical cosmological perturbations of the scalar fields has the same form as that of vacuum fluctuations due to squeezing of the quantum states that leave the horizon and then reenter it \cite{squeezed-state, Liddle} (while there also some studies that question the rigor of this formulation \cite{not-squeezed-state}). We take $\chi$s participating in $\chi\chi\,\rightarrow\,\phi\phi$ to be due to super-horizon modes of perturbation just after inflation so that they do not have enough time to thermalize, so $\tilde{f}^{(i)}$ does not have a thermal distribution i.e. we assume that the distribution of $\chi$ in phase space is similar to that of the perturbations of inflaton just after inflationary era as given in (\ref{t3a}). In other words we assume that the source of the $\chi$ fields in $\chi\chi\,\rightarrow\,\phi\phi$ processes is $\chi$s in the super-horizon modes of the vacuum fluctuations of $\chi$ (or those due to decays of inflaton in the super-horizon modes of its perturbations) after inflation. Note that the form of the power spectrum of the super-horizon modes (that leave and reenter the horizon) and their origin i.e. the quantum vacuum fluctuations of the fields in the inflationary era have the same form. Therefore the distribution function obtained in Appendix C for the quantum vacuum fluctuations has the same form as (\ref{t3a}) which is a (semi) classical distribution.

The distribution of $\phi$ particles in phase space at early times after the start of the process $\chi\chi\,\rightarrow\,\phi\phi$  may be determined from (\ref{t1}) after using (\ref{t3a}). For sake of simplicity we consider the initial times where $\tilde{f}^{(f)}(\vec{\tilde{p}}_j,\eta)\simeq\,0$, $j=3,4$. Then (\ref{t1}) becomes
\begin{eqnarray}
&&\frac{d\,\tilde{f}^{(f)}(\vec{\tilde{p}}_4,\eta)}{d\eta}\,\simeq\,
\tilde{n}_\chi^2\,\frac{\left[\Theta\left(|\vec{\tilde{p}}|_{max}-|\vec{\tilde{p}}|_{min}\right)-
\Theta\left(|\vec{\tilde{p}}|_{min}-|\vec{\tilde{p}}|_{max}\right)\right]}
{\left(|\vec{\tilde{p}}|_{max}-|\vec{\tilde{p}}|_{min}\right)^2}
\,{\cal B}(|\vec{\tilde{p}}_4|) \nonumber \\
&=&
\tilde{n}_\chi^2\,\frac{\left[\Theta\left(\sqrt{\vec{\tilde{k}}_{max}^2+\tilde{m}_\phi^2
-\tilde{m}_\chi^2}-\sqrt{\tilde{m}_\phi^2-\tilde{m}_\chi^2}|\right)-
\Theta\left(\sqrt{\tilde{m}_\phi^2-\tilde{m}_\chi^2}-\sqrt{\vec{\tilde{k}}_{max}^2+\tilde{m}_\phi^2
-\tilde{m}_\chi^2}\,\right)\right]}
{\left(\sqrt{\vec{\tilde{k}}_{max}^2+\tilde{m}_\phi^2-\tilde{m}_\chi^2}-\sqrt{\tilde{m}_\phi^2-\tilde{m}_\chi^2}\right)^2} \nonumber \\
&&~~~~~~~~\times\,\,{\cal B}(|\vec{\tilde{p}}_4|)\;,
\label{t2v}
\end{eqnarray}
where we have used $|\vec{\tilde{p}}|_{max}=\sqrt{|\vec{\tilde{k}}|_{max}^2+\tilde{m}_\phi^2-\tilde{m}_\chi^2}$, $|\vec{\tilde{p}}|_{min}=\sqrt{\tilde{m}_\phi^2-\tilde{m}_\chi^2}$.
Here
\begin{eqnarray}
&&{\cal B}(|\vec{\tilde{p}}_4|)
\,=\,
\frac{1}{128(2\pi)^7\tilde{E}_4}\int\int\int\,
\frac{d^3\vec{\tilde{p}}_1}{\tilde{E}_1}\frac{d^3\vec{\tilde{p}}_2}{\tilde{E}_2}\frac{d^3\vec{\tilde{p}}_3}{\tilde{E}_3}\,
\delta^{(4)}(\tilde{p}_1+\tilde{p}_2-\tilde{p}_3-\tilde{p}_4)
\,|\tilde{M}|^2\nonumber \\
&&\times\,\frac{\left[\Theta\left(|\vec{\tilde{p}}_1|-|\vec{\tilde{p}}|_{min}\right)-
\Theta\left(|\vec{\tilde{p}}_1|-|\vec{\tilde{p}}|_{max}\right)\right]\left[\Theta\left(|\vec{\tilde{p}}_2|-|\vec{\tilde{p}}|_{min}\right)-
\Theta\left(|\vec{\tilde{p}}_2|-|\vec{\tilde{p}}|_{max}\right)\right]}{\vec{\tilde{p}}_1^2 \,\vec{\tilde{p}}_2^2} \;.
\label{ab2xxa3}
\label{x1}
\end{eqnarray}
Note that the $|\vec{\tilde{k}}|_{max}$ dependence in (\ref{x1}) may be eliminated (in favor of $|\vec{\tilde{p}}_4|$ and a constant $|\alpha|_{max}$) by making use of
\begin{eqnarray}
&&\vec{\tilde{p}}_3=\tilde{m}_3\frac{\vec{\tilde{P}}}{\tilde{m}_3+\tilde{m}_4}+\vec{\tilde{k}}=\frac{\vec{\tilde{P}}}{2}+\vec{\tilde{k}}~,~~
\vec{\tilde{p}}_4=\tilde{m}_4\frac{\vec{\tilde{P}}}{\tilde{m}_3+\tilde{m}_4}-\vec{\tilde{k}}=\frac{\vec{\tilde{P}}}{2}-\vec{\tilde{k}}
\label{ab1a} \\
&&\vec{\tilde{P}}=\vec{\tilde{p}}_3+\vec{\tilde{p}}_4~,~~
\vec{\tilde{k}}=\frac{(\tilde{m}_2\vec{\tilde{p}}_3-\tilde{m}_1\vec{\tilde{p}}_4)}{\tilde{m}_3+\tilde{m}_4}=\frac{\vec{\tilde{p}}_3-\vec{\tilde{p}}_4}{2} \;.
\label{ab1b}
\end{eqnarray}
Eq.(\ref{ab1b}) implies that, for a fixed $\vec{\tilde{p}}_4$, there exist the largest number $|\alpha|_{max}$ that maximizes $|\vec{\tilde{k}}|$ with $\vec{\tilde{p}}_3=-|\alpha|_{max}\vec{\tilde{p}}_4$  i.e.
$|\vec{\tilde{k}}|_{max}=\frac{1}{2}\left(1+|\alpha|_{max}\right)|\vec{\tilde{p}}_4|$.

Eq.(\ref{x}) tells us that $|\vec{\tilde{k}}|_{max}$ goes to zero by time. The rate of the process in the reverse direction $\phi\phi\,\rightarrow\,\chi\chi$ is proportional to the the relative velocities of $\phi$s, so it is proportional to $|\vec{\tilde{k}}|$. These two phenomena together imply that the average value of $|\vec{\tilde{k}}|$ and its maximum value $|\vec{\tilde{k}}|_{max}$ approach zero by time. This verifies the assumption $\vec{\tilde{k}}^2\,\ll\,\tilde{m}_\phi^2-\tilde{m}_\chi^2$. Thus, eventually, $\tilde{f}^{(f)}$ becomes
\begin{eqnarray}
&&\frac{d\,\tilde{f}^{(f)}(\vec{\tilde{p}}_4,\eta)}{d\eta}\,\simeq\,
\tilde{n}_\chi^2\,\frac{\delta\left(\sqrt{\frac{1}{4}\vec{\tilde{p}}_4^2+\tilde{m}_\phi^2-\tilde{m}_\chi^2}-\sqrt{\tilde{m}_\phi^2-\tilde{m}_\chi^2}|\right)}
{\left(\sqrt{\frac{1}{4}\vec{\tilde{p}}_4^2+\tilde{m}_\phi^2-\tilde{m}_\chi^2}-\sqrt{\tilde{m}_\phi^2-\tilde{m}_\chi^2}\right)}
\,\,{\cal B}(|\vec{\tilde{p}}_4|)\;,
\label{t2vx}
\end{eqnarray}
where $\delta$ denotes Dirac delta function, and we have used $lim_{x\rightarrow\,y}\frac{\Theta(y)-\Theta(x)}{y-x}=\delta(x)$. Eq.(\ref{t2vx}) implies that the system reaches coherence about $|\vec{\tilde{p}}_4|=0$ by time which is one of the main properties of Bose-Einstein condensation.
Note that, although the equation (\ref{t2vx}) is an important indication for the formation Bose-Einstein condensation, it does not prove its formation. Eq. (\ref{t2vx}) is obtained from (\ref{t2v}) that is obtained at initial time when $\tilde{f}^{(f)}(\vec{\tilde{p}}_j,\eta)\ll\,1$ (when the condensate has not formed yet) while (\ref{t2vx}) is for much later times when $|\vec{\tilde{k}}|_{max}\,\rightarrow\,0$. In other words, what we have shown is not a proof but a hint towards formation of a condensate at later times.  Moreover, although, evolution towards $|\vec{\tilde{p}}_4|=0$ is an important indication towards formation of condensation it is not sufficient \cite{Semikoz}. In fact, essentially, (\ref{x}) also shows evolution towards $|\vec{\tilde{p}}_4|=0$. (\ref{t2vx}) reiterates and reinforces this result, and encourages for a more comprehensive study. Therefore we have only shown the tendency towards condensation rather than proving its formation. For a rigorous check of formation of condensation one must repeat the same steps for all times including the times  when the effect of Bose statistics can not be ignored (i.e. including the times when $\tilde{f}^{(f)}(\vec{\tilde{p}}_j,\eta)$ cannot be neglected on the right hand side of (\ref{t1})) and then solve the equation for $\tilde{f}^{(f)}(\vec{\tilde{p}}_4,\eta)$ and show that it has a delta function of the form of (\ref{t2vx}). This needs a separate study by its own in future. Another point worth to mention is that the processes $\phi\phi\,\rightarrow\,\chi\chi$ and $\chi\phi\,\rightarrow\,\chi\phi$ also take place. However, the rates of these processes are small at initial times as may be seen from Eq.(\ref{t1}) since we assume that initially we have only $\chi$ particles, so the number density of $\phi$ particles in phase space is small. Moreover, the process $\chi\phi\,\rightarrow\,\chi\phi$ is a number conserving process, so it is not expected to disturb condensation \cite{kinetic-theory}. Although $\phi\phi\,\rightarrow\,\chi\chi$ is a number changing process for $\phi$, there is an overall increase in the number density of $\phi$ particles when one considers both of $\chi\chi\,\rightarrow\,\phi\phi$ and $\phi\phi\,\rightarrow\,\chi\chi$, so the overall evolution, at least till the time of chemical equilibrium, is towards formation of condensation. To have a more comprehensive and rigorous picture of evolution of condensation, all these points must considered in detail in future studies.

The second condition for formation of Bose-Einstein condensation is the requirement of the overlap of the de Broglie wavelengths of the $\phi$ particles i.e. the condition
\begin{equation}
\frac{1}{|\vec{\tilde{p}}_4|}\,>\,\tilde{n}_\phi^{-\frac{1}{3}} \label{t4}
\end{equation}
(that guarantees the long range correlation of the system at late times)
is always satisfied for $|\vec{\tilde{k}}_{max}|\rightarrow\,0$. (Note that the co-moving number density $\tilde{n}$ is related to the number density by $\tilde{n}=a^3\,n$.) In the following paragraphs we show that the final condition for Bose-Einstein condensation (in cosmology), namely,
sufficiently fast production of $\phi$ particles so that the number density of $\phi$, $n_\phi$ may reach a sizable finite value in the presence of cosmological expansion is realized in the present model.

\subsection{Achieving finite number density for the condensation}

By definition the number density $\tilde{n}$ is related to $\tilde{f}$ by $\tilde{n}=\int\tilde{f}d^3\tilde{p}$. Hence after integrating (\ref{t1}) over $\vec{\tilde{p}}_4$, for initial times of the transition $\chi\chi\,\rightarrow\,\phi\phi$ (where $\tilde{f}_3$, $\tilde{f}_4$ on the right side of (\ref{t1}) may be neglected) one obtains
\begin{eqnarray}
\frac{d\,\tilde{n}_4(\eta)}{d\eta}&=&\frac{\tilde{v}}{2(2\pi)^3}\int\int\,
\tilde{f}_1\tilde{f}_2\,
d^3\vec{\tilde{p}}_1\,d^3\vec{\tilde{p}}_2\,\int\int\,d\tilde{\sigma}\;, \label{t1sx}
\end{eqnarray}
where differential of cross section $d\tilde{\sigma}$ is
\begin{eqnarray}
d\tilde{\sigma}&=&(2\pi)^4\,\delta^{(4)}(\tilde{p}_1+\tilde{p}_2-\tilde{p}_3-\tilde{p}_4)\frac{1}{4\tilde{E}_1\tilde{E}_2\tilde{v}}\,
\,|\tilde{M}|^2\nonumber\frac{d^3\vec{\tilde{p}}_3}{(2\pi)^32\tilde{E}_3} \frac{d^3\vec{\tilde{p}}_4}{(2\pi)^32\tilde{E}_4} \;. \label{t1sa}
\end{eqnarray}
In the case where the variation of $d\tilde{\sigma}$ with $\tilde{E}_1$, $\tilde{E}_2$ is small, the integral on the right hand side of (\ref{t1sx}) becomes $\int\int\,
\tilde{f}_1\tilde{f}_2\,
d^3\vec{\tilde{p}}_1\,d^3\vec{\tilde{p}}_2\,\int\int\,d\tilde{\sigma}$=$\tilde{\beta}\tilde{n}_1\tilde{n}_2\tilde{\sigma}$ with $\tilde{\beta}\,\sim\,1$. Thus, (\ref{t1sx}) may be expressed as
\begin{equation}
\frac{d\,\tilde{n}_4(\eta)}{d\eta}\,=\,\beta\tilde{n}_1\tilde{n}_2\tilde{\sigma}\,\tilde{v} \label{t1sc}
\end{equation}
where $\beta$ is a constant that corresponds to average effective depth of the collisions, $\tilde{\sigma}$ is the total cross-section
of the process, $\tilde{v}$ is the average relative velocity of two initial particles in the space defined by (\ref{t1ax}).

We have $\tilde{n}=\int\tilde{f}d^3\tilde{p}$=$\int\,f\,d^3\tilde{p}$=$a^3\,\int\,f\,d^3p=a^3\,n$ where we have used (\ref{s1}) and $d^3\tilde{p}=a^3\,d^3p$. Hence, (\ref{t1sc}) may be expressed as
\begin{equation}
\dot{n}_4+3H\,n_4\,=\,\beta\,n_1n_2\sigma\,v \;.\label{t1sd}
\end{equation}
Here
\begin{equation}
\sigma\,=\,a^2\,\tilde{\sigma}~,~~v\,=\,|\vec{v}|~,~~\vec{v}\,=\,a\frac{d\vec{r}}{dt}=\,\frac{d\vec{r}}{d\eta}\,=\,\vec{\tilde{v}} \;,\label{t1sdx}
\end{equation}
where we have used the fact that $\vec{v}$ is the (relative) peculiar velocity of the particles (in a cosmological context).
After using $n=\frac{C(t)}{a^3}$ one obtains
\begin{equation}
\dot{n}+3H\,n\,=\,\frac{\dot{C}}{a^3}~~~\mbox{where}~~n(t)=\frac{C(t)}{a^3(t)}
\label{t5}
\end{equation}
so, after comparing (\ref{t5}) and (\ref{t1sd}) for the process $\chi\chi\,\rightarrow\,\phi\phi$ one obtains
\begin{equation}
\frac{\dot{C}_\chi}{a^3}\,=\,-\frac{\dot{C}_\phi}{a^3}\,=\,-\beta\,n_\chi^2\,\sigma\,v \;.\label{t6}
\end{equation}
One may solve (\ref{t6}) to determine $C_\phi$, and hence $n_\phi$
once the dependence of $\sigma$ on $a(t)$ is determined \cite{Erdem}.
We take $\vec{\tilde{p}}^2\,\gg\,\tilde{m}_\chi^2$, $\vec{\tilde{k}}^2\,\ll\,\tilde{m}_\phi^2$, so (\ref{t1b}), after using (\ref{x}), may be approximated by
\begin{equation}
\tilde{M}\,\simeq\,\left(\frac{\tilde{\mu}}{\tilde{m}_\phi}\right)^2\frac{1}{1-\frac{\vec{\tilde{k}}^2}{\tilde{m}_\phi^2}\cos^2{\theta}} \;.
\label{t1bx}
\end{equation}
The cross section for $\chi\chi\,\rightarrow\,\phi\phi$ in the effective Minkowski space (for $\tilde{m}_\phi\,\gg\,\tilde{m}_\chi$) is given by
\begin{equation}
\tilde{\sigma}\,=\,
\frac{(2\pi)^4}{4\sqrt{(\tilde{p}_1.\tilde{p}_2)^2-\tilde{m}_\chi^4}}\int\int\,\delta^{(4)}(\tilde{p}_1+\tilde{p}_2-\tilde{p}_3-\tilde{p}_4)\,|M|^2
\frac{d^3\vec{\tilde{p}}_3}{\tilde{E}_3}\frac{d^3\vec{\tilde{p}}_4}{\tilde{E}_4}\,\simeq\,\left(\frac{\tilde{\mu}}{\tilde{m}_\phi}\right)^4
\frac{|\vec{\tilde{k}}|}{64\pi\,|\vec{\tilde{p}}|^2\tilde{m}_\phi} \;,\label{t8}
\end{equation}
where we have used the approximation
$\tilde{M}=\frac{\left(\frac{\tilde{\mu}}{\tilde{m}_\phi}\right)^2}{1-\frac{|\vec{\tilde{k}}|^2}{\tilde{m}_\phi^2}\cos^2{\theta}}
\simeq\,\left(\frac{\mu}{\tilde{m}_\phi}\right)^2$ since $\vec{\tilde{k}}^2\ll\,\tilde{m}_\phi^2$.

There are three different possible cases:\\
i)  $\frac{\dot{a}^2}{a^2}+\frac{\ddot{a}}{a}\,\leq\,m_\chi^2$ and $\frac{\dot{a}^2}{a^2}+\frac{\ddot{a}}{a}\,\geq\,0$,\\
ii)  $\frac{\dot{a}^2}{a^2}+\frac{\ddot{a}}{a}\,\leq\,m_\chi^2$ and $\frac{\dot{a}^2}{a^2}+\frac{\ddot{a}}{a}\,<\,0$,\\
iii) $\frac{\dot{a}^2}{a^2}+\frac{\ddot{a}}{a}\,>\,m_\chi^2$ \\
The case iii) above should be excluded to get rid of troublesome tachyons. Therefore the cases i) and ii) remain as the only safe choices. In the case of (\ref{t9a3}), i) implies that $\frac{\dot{a}^2}{a^2}+\frac{\ddot{a}}{a}\,=\,\xi^2(2-s)\,a^{-2s}\,\geq\,0$ i.e. $s\,\leq\,2$ while ii) implies that $\xi^2(2-s)\,a^{-2s)}\,<\,0$ i.e. $s\,>\,2$. Most of the simple physically relevant cosmological eras (namely, radiation dominated, matter dominated, and cosmological constant dominated eras) correspond to $s\,\leq\,2$ while the only physically interesting era for the case $s\,>\,2$ is a possible stiff matter dominated era where $s=3$. Therefore we consider the case i) here while we study the extreme case $|\frac{\dot{a}^2}{a^2}+\frac{\ddot{a}}{a}|\,\gg\,m_\phi^2$ of $s\,>\,2$ in the Appendix D to see the basic implications of ii).

For the case i) above we have $|\frac{\dot{a}^2}{a^2}+\frac{\ddot{a}}{a}|\,\ll\,m_\phi^2$ since $m_\chi^2\,\ll\,m_\phi^2$. This, in turn, implies that
$\sigma\,=\,a^2\tilde{\sigma}\simeq\,a^2\left(\frac{\tilde{\mu}}{\tilde{m}_\phi}\right)^4\frac{|\vec{\tilde{k}}|}{64\pi\,\vec{\tilde{p}}^2\tilde{m}_\phi}\,\propto\,a$ since $\tilde{m}_\phi\,\simeq\,a\,m_\phi$, $\tilde{\mu}\,\simeq\,a\,\mu$,  and $|\vec{\tilde{p}}|$, $|\vec{\tilde{k}}|$ are independent of redshift by (\ref{t1y}).
In this case (\ref{t6}) results in
\begin{eqnarray}
&&\frac{\dot{C}_\chi}{a^3}\,=\,-\beta\,\left(\frac{C_\chi}{a^3}\right)^2\,\sigma_0\,v_0\,a \label{t9a1} \\
&&\Rightarrow~~\frac{dC_\chi}{C^2}\,=\,- \frac{1}{\xi}\beta\,\sigma_0\,v_0\,a^{s-3}\,da \;,\label{t9a2}
\end{eqnarray}
where we have taken
\begin{equation}
H\,=\,\xi\,a^{-s} \;,\label{t9a3x}
\end{equation}
which includes all simple interesting cases e.g. radiation, matter, stiff matter, cosmological constant dominated universes.
After integrating out (\ref{t9a2}) we find
\begin{equation}
C_\chi\,=\,\frac{C_1}{\frac{C_1\beta\sigma_0v_0}{(s-2)\xi}\left(\,a^{s-2}-a_1^{s-2}\right)+1} \;,\label{t9a4}
\end{equation}
where $C_1=C_\chi(t_1)$ is the value of $C_\chi$ at the start of the conversion of $\chi$s to $\phi$s.
After the use of (\ref{t6}) and (\ref{t9a4}) one obtains
\begin{equation}
C_\phi\,=\,\int_{t_1}^t\frac{dC_\phi}{dt}dt\,=\,-\int_{t_1}^t\frac{dC_\chi}{dt}dt\,=\,C_1\,-\,
\frac{C_1}{\frac{C_1\beta\sigma_0v_0}{(s-2)\xi}\left(\,a^{s-2}-a_1^{s-2}\right)+1}\;.
\label{t9a6}
\end{equation}
We see that $n_\phi=\frac{C_\phi}{a^3}$ may reach finite values if $C_\phi\,\propto\,a^r$ where $r\,>\,0$ since $C_\phi$ is initially zero. At initial times where $t\simeq\,t_1$, (\ref{t9a6}) may be approximated by
\begin{equation}
C_\phi\,\simeq\,
\frac{C_1\beta\sigma_0v_0}{(s-2)\xi}\left(\,a^{s-2}-a_1^{s-2}\right) \;,
\label{t9a6a}
\end{equation}
which may be expressed as
\begin{eqnarray}
&&C_\phi\,\simeq\,
\frac{C_1\beta\sigma_0v_0}{(|s-2|)\xi}\,a^{|s-2|}\left[\,1-\left(\frac{a_1}{a}\right)^{|s-2|}\right]~~~~\mbox{for}~~s-2\,>\,0  \;,
\label{t9a6aa} \\
&&C_\phi\,\simeq\,
\frac{C_1\beta\sigma_0v_0}{(|s-2|)\xi}a_1^{-|s-2|}\left[\,1-\left(\frac{a_1}{a}\right)^{|s-2|}\right]~~~~\mbox{for}~~s-2\,<\,0  \;,
\label{t9a6ab}
\end{eqnarray}
where (\ref{t9a6ab}) is the tachyon free case, so is the physically safe option.
(\ref{t9a6aa}) and (\ref{t9a6ab}) imply that $C_\phi$ initially has higher values for $s\,<\,2$ while $C_\phi$ grows faster for $s\,>\,2$.
 At late times (\ref{t9a6}) may be approximated by
\begin{eqnarray}
C_\phi\,\sim\,C_1\,-\,
\frac{(|s-2|)\xi}{\beta\sigma_0v_0}\,a^{-|s-2|}~~~~\mbox{for}~~s-2\,>\,0~\mbox{and}~\frac{C_1\beta\sigma_0v_0}{(|s-2|)\xi}\,a^{|s-2|}\,\gg\,1  \;,
\label{t9a6ac}
\end{eqnarray}
and
\begin{eqnarray}
C_\phi\,\sim\,C_1\,-\,
\frac{(|s-2|)\xi}{\beta\sigma_0v_0}\,a_1^{|s-2|}~~~~\mbox{for}~~s-2\,<\,0~\mbox{and}~\frac{C_1\beta\sigma_0v_0}{(|s-2|)\xi}\,a_1^{-|s-2|}\,\gg\,1  \;,
\label{t9a6ad}
\end{eqnarray}
where (\ref{t9a6ad}) is the tachyon free case, so is the physically safe option.
(\ref{t9a6ac}) and (\ref{t9a6ad}) imply that if the processes $\chi\chi\,\rightarrow\,\phi\phi$ continue till very late times then $C_\phi$ may reach its maximum value $C_1$ for $s\,>\,2$ while it will be smaller than that value in the case $s\,<\,2$. However the conclusions for late times here are not wholly reliable since we have neglected the effect of statics and the effect of the processes $\phi\phi\,\rightarrow\,\chi\chi$ which can not be neglected while they can be neglected at initial times.  Eqs. (\ref{t9a6ac}) and (\ref{t9a6ad}) are reliable only if the number density of $\phi$s has not reached a large value at late times yet. One needs to do a more detailed study in future to see the situation at late times clearly.

A final comment is in order here: In the light of (\ref{t8}), it is useful to check the range of validity of the condition (\ref{t1x}) in such perturbative calculations (where $\left(\frac{\tilde{\mu}}{\tilde{m}_\phi}\right)\,<\,1$) since smaller $\frac{\tilde{\mu}}{\tilde{m}_\phi}$ makes
$\sigma$ smaller. However, as is evident from (\ref{t8}), $\frac{\tilde{\mu}}{\tilde{m}_\phi}$ is not the only quantity that determines the magnitude of
$\tilde{\sigma}$. Moreover, the relevant quantity in (\ref{t1x}) is $n\sigma$ rather than $\sigma$ alone. Therefore, even for perturbative regime there is a considerable parameter space where such an effective Minkowski space formulation holds. For example, one may identify $\phi$ by dark matter and let $\frac{\tilde{\mu}}{\tilde{m}_\phi}=0.1$, $\frac{|\vec{\tilde{k}}|}{|\vec{\tilde{p}}|}=0.01$, $|\vec{\tilde{p}}|\sim\,\tilde{m}_\phi\,c$; then $n_0\,\sim\,\frac{10^{-3}\,eV\,cm^{-3}}{\tilde{m}_\phi\,c^2}$, so by (\ref{t8}), $n_0\tilde{\sigma}_0\simeq\,\frac{10^{-3}\,eV\,cm^{-3}}{\tilde{m}_\phi\,c^2}\left(\frac{\tilde{\mu}}{\tilde{m}_\phi}\right)^4\left(\frac{|\vec{\tilde{k}}|}{|\vec{\tilde{p}}|}\right)
\left(\frac{\hbar\,c}{\tilde{m}_\phi\,c^2}\right)^2$$\sim$$10^{-19}\times\left(\frac{eV}{\tilde{m}_\phi}\right)^3\,cm^{-1}$ which satisfies  $\frac{H_0}{n_0\sigma_0\,v}\simeq\,\frac{10^{-28}cm^{-1}}{n_0\sigma_0(v/c)}\,\ll\,1$ provided that $\frac{v}{c}\sim\,1$ and $\tilde{m}_\phi\,\ll\,10^3\,eV$ (including the phenomenologically interesting case of ultra light dark matter). Note that here we  cannot take $|\vec{\tilde{k}}|\,\simeq\,0$ since that would correspond to full Bose-Einstein condensation and in that case the whole system behaves as a single quantity, so the usual formulation of scattering in terms of single particles in quantum field theory does not work. In fact, our aim in this paper is to show how curved space effects promote formation of Bose-Einstein condensation in a model that contains $\phi^2\chi$ type of interaction terms rather than to show that an effective Minkowski space formulation works in all cases in cosmology. We hope that this formulation and this study may give an additional insight in understanding formation of Bose-Einstein condensation in cosmology.

\section{Conclusion}

 In this study $\chi\chi\,\rightarrow\,\phi\phi$ processes that are induced by $\phi^2\chi$ terms in the Lagrangian in a flat Robertson-Walker background are considered. Some conditions on the rate of $\chi\chi\,\rightarrow\,\phi\phi$ processes and the magnitude of the curvature with respect to the particle masses are imposed so that the spacetime is effectively Minkowskian at the time scale of each individual process to make the analysis simpler. In the preceding sections it is shown that all conditions of the pre-condensation of $\phi$ particles are satisfied in this setup. A more comprehensive study of the effect of Bose-Einstein statics through numerical calculations must be done in future to show the same behaviour in later stages of evolution of Bose-Einstein condensation (except the final stage where condensation is wholly achieved since at that stage the usual scattering theory of quantum field theory that is employed in this study cannot be used). This setup has a promising prospect that the same trilinear coupling $\phi^2\chi$ induces $\phi\phi\,\rightarrow\,\phi\phi$ processes that may provide the conditions for thermalization of $\phi$ particles and correspond to an effective $\lambda\phi^4$ term in the Lagrangian. This point deserves a separate study by its own in future. This scheme, in addition to providing a microscopic description of BEC at the particle physics level, has an additional advantage of providing redshift dependent $\lambda$, so providing a richer phenomenological prospect. This property of the scheme will be more evident in future studies where the approximate analysis in this study is extended to a more general framework.

\begin{acknowledgments}
We would like to thank Professor Masahiro Morikawa for reading the manuscript and for his valuable comments.

This paper is financially supported by {\it The Scientific and Technical Research Council of Turkey (T{\"{U}}BITAK)} under the project 117F296 in the context of the COST action {\it CA 16104 "GWverse"}
\end{acknowledgments}

\appendix

\section{Condition (\ref{t1x}) in the case of radiation, matter, cosmological constant, and stiff matter dominated eras}

If we let
\begin{equation}
H\,=\,\xi\,a^{-s} \;,\label{t9a3}
\end{equation}
which includes all simple interesting cases e.g. radiation, matter, stiff matter, cosmological constant dominated universes, then (\ref{t1x}) becomes
\begin{equation}
\left|\frac{2a^2H\left[m_\chi^2-(s^2-3s+2)H^2\right]}{n_\chi\beta\sigma\,v\,a^2\left(m_\chi^2-(2-s)H^2\right)}\right|
\,=\,\left|\frac{2H}{n_\chi\beta\sigma\,v}\left(1-\frac{s(s-2)H^2}{m_\chi^2+(s-2)H^2}
\right)\right|\ll\,1\;. \label{t1xxx}
\end{equation}

Essentially, there two characteristic cases that satisfy (\ref{t1xxx}): \\
${\bf i}$- $|\left(1-\frac{s(s-2)H^2}{m_\chi^2+(s-2)H^2}\right)|\,\lesssim\,\mathcal{O}(1)$ (i.e.
$s\simeq\,0$ or $s\simeq\,2$ provided that $\frac{m_\chi^2}{H^2}$ is not close to $2-s$ or $\frac{m_\chi^2}{(s-2)H^2}+1\,\gg\,s$). In this case it is enough to have $|\frac{2H}{n_\chi\beta\sigma\,v}|\,\ll\,1$ to satisfy (\ref{t1xxx}). Note that $|\frac{H}{n_\chi\beta\sigma\,v}|\,<\,1$ should already be satisfied to enable the process to take place. \\
${\bf ii}$- $|\frac{2H}{n_\chi\beta\sigma\,v}|\,\lesssim\,\mathcal{O}(1)$. In that case it is enough to have    $|\left(1-\frac{s(s-2)H^2}{m_\chi^2+(s-2)H^2}\right)|\,\ll\,1$ i.e. to set $\frac{m_\chi^2}{H^2}\,\simeq\,(s-1)(s-2)$ which also implies that $s\,>\,2$ or $s\,<\,1$ (in addition to $|\frac{m_\chi^2}{H^2}|\,\simeq\,|(s-1)(s-2)|$).

The condition (\ref{t1xxx}) is satisfied for a considerable range of parameters. For example, the case i) above may be realized in the radiation dominated era ($s=2$) well after its start (to make $|\frac{2H}{n_\chi\beta\sigma\,v}|\,\ll\,1$ applicable) independent of the value of $m_\chi$. The case i) is also satisfied for the current epoch of accelerated cosmic expansion (where $s\,\simeq\,0$) independent of the value of $m_\chi$ provided that $\frac{H_0}{n_0\beta\sigma_0\,v}\simeq\,\frac{H_0}{n_0\sigma_0\,v}\simeq\,\frac{10^{-26}m^{-1}}{n_0\sigma_0(v/c)}\,\ll\,1$. For example, for $n_0\,\gg\,10^{14}(meter)^{-3}$ with $\sigma_0\sim\,(\frac{c}{v})10^{-40}\,(meter)^2$, and for $n_0\,\gg\,10^{7}(meter)^{-3}$ with $\sigma_0\sim\,(\frac{c}{v})10^{-33}\,(meter)^2$ (where $c$ denotes the speed of light) $\frac{H_0}{n_0\beta\sigma_0\,v}\,\ll\,1$ is easily satisfied.  Note that the order of electromagnetic interaction and the weak interaction cross sections are $10^{-32}\,(meter)^2$
and $10^{-40}\,(meter)^2$, respectively, and the number density of photons and baryons at present are $>\,10^8\,(meter)^{-3}$ and $\sim\,10^{-1}\,(meter)^{-3}$, respectively. It is evident that, by adjusting the values of cross sections and number densities accordingly, one may cover a much wider parameter space than those in the above examples. Therefore, this method can be used for a sufficiently wide range of models. The case ii) may be satisfied in a possible stiff matter dominated era ($s=3$) after inflation or in the current accelerated expansion era, $s\neq\,0\sim\,0$ provided that $m_\chi^2\sim\,2H_0^2$.

\section{Formal derivation of Equation (\ref{e2aaaxx})}

A field $\tilde{\chi}$ may be expanded in its Fourier modes as
\begin{equation}
\tilde{\chi}(\vec{r},\eta)\,=\,\frac{1}{\sqrt{2}}\int\,\frac{d^3\tilde{p}}{(2\pi)^\frac{3}{2}}\left[a_p^-\,v_p^*(\eta)e^{i\vec{\tilde{p}}.\vec{r}}
\,+\,a_p^+\,v_p(\eta)e^{-i\vec{\tilde{p}}.\vec{r}}\right] \;,
\label{e2aaa}
\end{equation}
where $\vec{r}=(\tilde{x}_1,\tilde{x}_2,\tilde{x}_3)$, $a_p$ are the expansion coefficients that are identified by annihilation operators after quantization, $v_p$ are the basic normalized solutions of the equation of motion of $\tilde{\chi}$. In the time interval between two particle physics processes (such as those in Figure \ref{fig:1}) one may consider $\tilde{\chi}$ particles to be free particles, so they satisfy
\begin{equation}
v_p^{\prime\prime}\,+\,\omega_p^2(\eta)\,v_p\,=\,0~~~\mbox{where}~~\omega_p\,=\,\sqrt{|\vec{\tilde{p}}|^2+\tilde{m}_\chi^2} ~~,~~~
v_p^\prime\,v_p^*-v_p\,v_p^{*\,\prime}\,=\,2i \;. \label{e2aab1}
\end{equation}
We require that the variations in $\tilde{m}_\chi^2$ and $\tilde{m}_\phi^2$ are small (that are already insured by (\ref{t1x})). This, in turn, implies that the variation in $\omega_p^2$ of (\ref{e2aab1}) with time is small since $\vec{\tilde{p}}$ does not depend on time. Therefore we may adopt WKB approximation \cite{QFTC} so that $v_p$ in the time interval between $\eta_i$ and $\eta_{i+1}$ is given by
\begin{equation}
v_p(\eta)\,=\,\frac{1}{\sqrt{W_p(\eta)}}exp{\left(i\int_{\eta_i}^\eta\,W_p(\eta)\,d\eta\right)} \label{e2c1}
\end{equation}
where $W_p$, by (\ref{e2aab1}), satisfies
\begin{equation}
W_p^2\,=\,\omega_p^2-\frac{1}{2}\left[\frac{W_p^{\prime\prime}}{W_p}-\frac{3}{2}\left(\frac{W_p^\prime}{W_p}\right)^2\right] \;.\label{e2c2}
\end{equation}
(\ref{e2c2}) may be solved approximately, for slowly varying $\omega_p$, as
\begin{equation}
^{(0)}W_p=\omega_p~,~~^{(2)}W_p=\omega_p\left(1-\frac{\omega_p^{\prime\prime}}{4\omega_p^3}+\frac{3\omega_p^{\prime\;2}}{8\omega_p^4}\right)~,~~\mbox{etc.}
\label{e2c3}
\end{equation}
where $^{(0)}W_p$ stands for the zeroth order approximation in $\frac{\Delta\,\omega_p}{\omega_p}$, and $^{(2)}W_p$ stands for the second order approximation in $\frac{\Delta\,\omega_p}{\omega_p}$ and is obtained by substituting  $^{(0)}W_p$ on its left hand side of (\ref{e2c2}) and then Taylor expanding the square root for slowly varying $\omega_p$. Higher order approximations are obtained in a similar manner.
Adopting WKB approximation for slowly varying $\omega_p$, and hence Eq.(\ref{e2c3}) is called adiabatic approximation \cite{Bunch,QFTC1}.

Adiabaticity condition simplifies the formulation and is useful to determine the approximate form of the mode function. But it does not make the mode function  Minkowskian by itself. When we also impose the condition
Eq.(\ref{t1x}) then we obtain
$^{(0)}W_p\,=\,\omega_p\,\simeq\,constant$ in the zeroth order approximation. Hence the space becomes approximately Minkowskian in each interval between $\eta_i$ and $\eta_{i+1}$ .  In this case $\omega_p$ does not only vary slowly but, in fact, it varies extremely slowly in each time interval by (\ref{t1x}). One may show that (\ref{t1x}) results in $^{(2)}W_p\,\simeq\,^{(0)}W_p\,=\,\omega_p\,\simeq\,constant$. This may be seen as follows: $\omega_p\,=\,\sqrt{|\vec{\tilde{p}}|^2+\tilde{m}_\chi^2}$ implies that
\begin{equation}
\frac{\omega_p^{\prime}}{\omega_p^2}\,=\,\frac{a}{\omega_p^2}\frac{d\omega_p}{dt}\,=\,\left(\frac{a}{2\omega_p^3}\right)\frac{d\tilde{m}_\chi^2}{dt}
\,=\,\left(\frac{a}{2\omega_p}\right)\left(\frac{\tilde{m}_\chi}{\omega_p}\right)^2(\Delta\,t)^{-1}
\left[\frac{\left(\frac{d\tilde{m}_\chi^2}{dt}\right)(\Delta\,t)}{\tilde{m}_\chi^2}\right]\,\simeq\,0
\label{rx1}
\end{equation}
where $\Delta\,t\,\sim\,\left(\frac{1}{n_\chi\beta\sigma\,v}\right)$. By (\ref{t1x}), the last term in the square brackets in (\ref{rx1}) is much smaller than one, so in order to show that (\ref{rx1}) is almost zero the remaining terms in (\ref{rx1}) must not be much larger than one. We observe that $\left(\frac{\tilde{m}_\chi}{\omega_p}\right)^2\,<\,1$. For reasonable values of the parameters $\left(\frac{a}{\omega_p}\right)(\Delta\,t)^{-1}\sim\left(\frac{a\,n_\chi\beta\sigma\,v}{\omega_p}\right)$ is much smaller than 1. For example, for the values of parameters discussed after the equation (\ref{t1xxx}) i.e. for
$\hbar\,n_\chi\beta\sigma\,v\,\gg\,\left(6.5\,\times\,10^{-16}\,eV.sec\,\times\,10^{-26}\,sec^{-1}\right)\,\simeq\, 10^{-41}\,eV$ we have  $\left(\frac{a}{\omega_p}\right)(\Delta\,t)^{-1}\sim\left(\frac{a\,n_\chi\beta\sigma\,v}{\omega_p}\right)\,\ll\,1$ provided that $\hbar\,\omega_p$ is much greater than $10^{-41}\,eV$.  Next we show that one may take $\frac{\omega_p^{\prime\prime}}{4\omega_p^3}\,\simeq\,0$ for reasonable values of the parameters. For simplicity, we consider the case in Appendix A i.e. the case $H=\xi\,a^{-s}$ (which includes all simple interesting cases e.g. radiation, matter, stiff matter, cosmological constant dominated universes). After using (\ref{t9a3}), and $\frac{d^2\omega_p^2}{dt^2}\,=\,\frac{d^2\tilde{m}_\chi^2}{dt^2}$, and
\begin{equation}
\frac{d^2\omega_p^2}{dt^2}\,=\,2\left(\frac{d\omega_p}{dt}\right)^2+2\omega\frac{d^2\omega_p}{dt^2},~~\mbox{and}~ \frac{d^2\tilde{m}_\chi^2}{dt^2}\,=\,2\left(\frac{d\tilde{m}_\chi}{dt}\right)^2+2\tilde{m}_\chi\frac{d^2\tilde{m}_\chi}{dt^2} \label{q}
\end{equation}
we observe that
\begin{eqnarray}
&&\frac{d\tilde{m}_\chi}{dt}\,=\,\frac{1}{2\tilde{m}_\chi}\frac{d\tilde{m}_\chi^2}{dt}\,=\,H\tilde{m}_\chi\,+\,\frac{a^2s(2-s)H^3}{\tilde{m}_\chi},
\nonumber \\
&&\frac{d~^2\tilde{m}_\chi}{dt^2}\,=\,
(1-s)\,H^2\tilde{m}_\chi\,+\,\frac{a^2s(2-s)(2-3s)H^4}{\tilde{m}_\chi}\,-\,\frac{a^4s^2(2-s)^2H^6}{\tilde{m}_\chi^3}\;. \label{rx2}
\end{eqnarray}
Therefore we have
\begin{eqnarray}
&&\frac{\omega_p^{\prime\prime}}{4\omega_p^3}\,=\,\frac{a}{4\omega_p^3}\frac{d\left(a\frac{d\omega_p}{dt}\right)}{dt}
\,=\,
\frac{a^2H\tilde{m}_\chi}{4\omega_p^4}\frac{d\tilde{m}_\chi}{dt}
\,+\,\frac{a^2}{4\omega_p^4}\left(1-\frac{\tilde{m}_\chi^2}{\omega_p^2}\right)\left(\frac{d\tilde{m}_\chi}{dt}\right)^2
\nonumber \\
&&+\,\frac{a^2\tilde{m}_\chi}{4\omega_p^4}\left[(1-s)\,H^2\tilde{m}_\chi\,+\,\frac{a^2s(2-s)(2-3s)H^4}{\tilde{m}_\chi}\,-\,\frac{a^4s^2(2-s)^2H^6}{\tilde{m}_\chi^3}\right]
\nonumber \\
&&=\,
\left(\frac{a}{\omega_p}(\Delta\,t)^{-1}\right)^2\left(\frac{\tilde{m}_\chi}{\omega_p}\right)^2
\{\frac{1}{8}H\Delta\,t\,
\left[\frac{\left(\frac{d\tilde{m}_\chi^2}{dt}\right)(\Delta\,t)}{\tilde{m}_\chi^2}\right]
\,+\,\frac{1}{16}
\left(1-\frac{\tilde{m}_\chi^2}{\omega_p^2}\right)\left[\frac{\left(\frac{d\tilde{m}_\chi^2}{dt}\right)(\Delta\,t)}{\tilde{m}_\chi^2}\right]^2
\nonumber \\
&&+\,\frac{1}{4}(H\Delta\,t)^2\left[
(1-s)\,\,+\,a^2s(2-s)(2-3s)\frac{H^2}{\tilde{m}_\chi^2}\,-\,a^4s^2(2-s)^2\left(\frac{H}{\tilde{m}_\chi}\right)^4\right]\}\,\simeq\,0\;,
\label{rx3}
\end{eqnarray}
provided that (\ref{t1x}) is satisfied. After using (\ref{rx1}) and (\ref{rx2}) one observes that (\ref{rx3}) too gives negligible contribution to (\ref{e2c3}) for reasonable values of the parameters provided that either $|H\Delta\,t|$ or $|\frac{H}{\tilde{m}_\chi}|$ are not much greater than 1.

The above analysis shows that, when (\ref{t1x}) is satisfied, $\omega_p=\mbox{constant}$ is a good approximation in each time interval $\eta_i\,<\,\eta\,<\,\eta_{i+1}$, so in each interval we may take $v_p\,\simeq\,v_p^{(i)}$ where
\begin{equation}
v_p^{(i)}(\eta)\,=\,\frac{1}{\sqrt{\omega_p^{(i)}}}exp{\left(i\omega_p^{(i)}(\eta-\eta_i)\right)}~~\mbox{where}~~\;
\omega_p^{(i)}=\omega_p(\eta_i)~,~~\eta_i\,<\,\eta\,<\,\eta_{i+1} \;. \label{e2c4}
\end{equation}
Hence (\ref{e2aaa}) may be expressed as
\begin{eqnarray}
\tilde{\chi}^{(i)}(\vec{r},\eta)\,\simeq\,
\int\,
\frac{d^3\tilde{p}}{(2\pi)^\frac{3}{2}\sqrt{2\omega_p^{(i)}}}\left[a_p^{(i)\,-}\,
e^{i\left(\vec{\tilde{p}}.\vec{r}-\omega_p^{(i)}(\eta-\eta_i)\right)}
\,+\,a_p^{(i)\,+}
\,e^{i\left(-\vec{\tilde{p}}.\vec{r}+\omega_p^{(i)}(\eta-\eta_i)\right)}\right]\label{e2aaax} \\
\eta_i\,<\,\eta\,<\,\eta_{i+1}\;, \nonumber
\end{eqnarray}
where $^{(i)}$ refers to the $i$th time interval between the $i$th and $(i+1)$th processes.

A comment is in order at this point. In the case $m_i^2a^2\,<\,\frac{a^{\prime\prime}}{a}$ in (\ref{e2aa}) (which corresponds to $m_i^2\,<\,(2-s)H^2$ in the case of (\ref{t9a3})) the effective mass $\tilde{m}_i$ becomes tachyonic. However $\frac{a^{\prime\prime}}{a}$ gets sufficiently small by time so that the particle masses become real after some time for all physically relevant cases (e.g. as given in (\ref{t9a3})) except in the case where strictly $s=0$. Therefore this is not a true problem in general for the physically interesting cases because either the mass becomes real after a finite time for $s\,>\,0$ or it can not interact with other particles (so, making the tachyonic state harmless)for $s\,\leq\,0$ due to fast expansion rate. However a tachyonic state can not be dealt within this formulation because the would-be ground state (e.g. $\chi=0$, $\phi=0$) will not be the ground state anymore, making the perturbation expansion about the ground state inapplicable. In other words this formulation is not applicable to the case, $m_i^2a^2\,<\,\frac{a^{\prime\prime}}{a}$ (which corresponds to $m_i^2\,<\,(2-s)H^2$ in the case of (\ref{t9a3})). Therefore the case of $s\,\geq\,2$ (e.g. of radiation and stiff matter) is safe in this regard while, in the the case of $s\,<\,2$ (e.g. cosmological constant, matter, radiation), $H^2$ should be sufficiently small compared to $m_\chi^2$ so that the problem of tachyons do not emerge. After combining this constraint with those discussed after (\ref{t1xxx}) one notices that there is still a significant relevant available parameter space left. The conclusions obtained after (\ref{t1xxx}) remain intact for radiation and stiff matter dominated eras, and the conclusions obtained for the current accelerated expansion era still hold provided that $m_\chi$ is not smaller than $\sim\,H_0\hbar\,\sim\,10^{-33}$ eV.

\section{The distribution function for vacuum fluctuations in homogenous and isotropic Spaces}

Because $\chi$ and $\phi$ in the initial times of the $\chi\chi\,\rightarrow\,\phi\phi$ transition (and in the Bose-Einstein condensation regime since the whole system acts as a single particle) the distribution of $\chi$ and $\phi$ may be effectively described by that of one particle states since they are not highly correlated yet.
Therefore number density (that is incorrectly identified as probability density in quantum mechanics \cite{Weinberg2}) of $\chi$s is proportional with its vacuum fluctuations $\tilde{n}_\chi^{(q)}$
\begin{equation}
\tilde{n}_\chi\,\propto\,\tilde{n}_\chi^{(q)}\,=\,lim_{\tilde{x}\rightarrow\,\tilde{y}}<0|2i\frac{\partial\tilde{\chi}(\tilde{x})}{\partial\,\eta}\tilde{\chi}(\tilde{y})|0> \;, \label{t3ya}
\end{equation}
(since in inflationary models the energy density perturbations are identified to be due to vacuum fluctuations)
where $\tilde{x}=(\eta_1,\vec{\tilde{r}}_1)$, $\tilde{y}=(\eta_2,\vec{\tilde{r}}_2)$, and $lim_{\tilde{x}\rightarrow\,\tilde{y}}$ denotes taking $\tilde{x}$ and $\tilde{y}$ being extremely close and averaging over that region rather than taking $\tilde{x}=\tilde{y}$ in order to avoid divergence of vacuum fluctuations \cite{BD},  and
\begin{eqnarray}
<2i\frac{\partial\tilde{\chi}(\tilde{x})}{\partial\,\eta}\tilde{\chi}(\tilde{y})>
&=&<0|2i\frac{\partial\tilde{\chi}(\tilde{x})}{\partial\,\eta}\tilde{\chi}(\tilde{y})|0>  \nonumber \\
&=&<0|\int\frac{d^3\tilde{q}}{(2\pi)^3}\,
\exp{\{i\left[\vec{\tilde{q}}.(\vec{\tilde{r}}_1-\vec{\tilde{r}}_2)-\omega_q(\eta_1-\eta_2)\right]\}}\,|0> \;. \label{t3y}
\end{eqnarray}
(Here we have used (\ref{e2aaax}) and $\eta$ refers to any $i$'th interval in (\ref{e2aaax})). On the other hand
\begin{eqnarray}
<\tilde{\chi}(\tilde{x})\tilde{\chi}(\tilde{y})>
&=&<0|\tilde{\chi}(\tilde{x})\tilde{\chi}(\tilde{y})|0> \nonumber \\
&=&<0|\int\frac{d^3\tilde{q}}{2(2\pi)^3\omega_q}\,\exp{\{i[\vec{\tilde{q}}.(\vec{\tilde{r}}_1-\vec{\tilde{r}}_2)-\omega_q(\eta_1-\eta_2)]\}}
\,|0> \;. \label{t3z}
\end{eqnarray}
Therefore
\begin{equation}
\tilde{f}^{(i)}(\vec{\tilde{p}},\eta)\,=\,\tilde{n}_{p\chi}(|\vec{\tilde{p}}|,\eta)\,=\,\omega_{p\chi}\,{\cal P}_\chi(|\vec{\tilde{p}}|,\eta)\,\simeq\,|\vec{\tilde{p}}|\;{\cal P}_\chi(|\vec{\tilde{p}}|,\eta) \;, \label{t3z1}
\end{equation}
where $\tilde{n}_{p\chi}$ and ${\cal P}_\chi(|\vec{\tilde{p}}|)$ are defined by \cite{Liddle}
\begin{eqnarray}
&&<2i\frac{\partial\tilde{\chi}(\tilde{x})}{\partial\,\eta}\tilde{\chi}(\tilde{y})>\,=\,\int\,d^3\tilde{q}\;
e^{i\vec{\tilde{q}}.(\vec{\tilde{r}}_1-\vec{\tilde{r}}_2)}\; \tilde{n}_{q\chi}(|\vec{\tilde{q}}|,\eta) \,\label{t3z2} \\
&&<\tilde{\chi}(\tilde{x})\tilde{\chi}(\tilde{y})>
\,=\,\int\,d^3\tilde{q}\;e^{i\vec{\tilde{q}}.(\vec{\tilde{r}}_1-\vec{\tilde{r}}_2)}{\cal P}_\chi(|\vec{\tilde{q}}|,\eta) \;.\label{t3z3}
\end{eqnarray}

In the case of scale invariance the spectrum ${\cal P}_\chi(|\vec{\tilde{p}}|,\eta)$ satisfies ${\cal P}_\chi(|\vec{\tilde{p}}|,\eta)\,\propto\,\frac{1}{|\vec{\tilde{p}}|^3}$ \cite{Weinberg} where it gains a $\eta$ dependence since it re-entered into horizon. This together with (\ref{t3z1}) results in a form similar to that of (\ref{t3a}).

\section{The evolution of the number densities in the case of $\frac{\dot{a}^2}{a^2}+\frac{\ddot{a}}{a}\,<\,0$ with $|\frac{\dot{a}^2}{a^2}+\frac{\ddot{a}}{a}|\,\gg\,m_\phi^2$}

 This case i.e. $|\frac{\dot{a}^2}{a^2}+\frac{\ddot{a}}{a}|$=$|\dot{H}+2H^2|\,\gg\,m_\phi^2$ does not seem to be relevant to a Bose-Einstein condensate dark matter $\phi$ since this case implies $|\dot{H}+2H^2_0|\,\sim\,H_0^2\,\sim\,(10^{-31}eV)^2\gg\,m_\phi^2$ i.e. the mass of the dark matter at present should much smaller than $10^{-33}eV$ which seems to be excluded by observations \cite{ULDM-mass}. However this case may be relevant for dark energy (provided the contribution of the potential term dominates) and for scalars in early universe.

In this case
\begin{equation}
\tilde{m}^2\,\simeq\,-a^2\left(\dot{H}+2H^2\right) \,=\,\xi^2(s-2)\,a^{2(1-s)}\,>\,0~~\mbox{i.e.}~s\,>\,2\label{t10}
\end{equation}
where we have taken $H\,=\,\xi\,a^{-s}$. Therefore
\begin{equation}
\tilde{m}^2\,\propto\,a^{2(1-s)}\,;~~\tilde{\mu}\,\propto\,a\,
~;~~~\sigma\,=\,a^2\tilde{\sigma}\,
\simeq\,a^2\left(\frac{\tilde{\mu}}{\tilde{m}_\phi}\right)^4\frac{|\vec{\tilde{k}}|}{64\pi\,\vec{\tilde{p}}^2\tilde{m}_\phi}\,\propto\,a^{5s+1} \;. \label{t11}
\end{equation}
(It is observed from (\ref{t11}) that $\frac{2H}{n\sigma\,v}=\frac{2H_0}{n_0\sigma_0\,v_0}\,a^{-6s+2}$, so (\ref{t1xxx}) (i.e. (\ref{t1x})) is always satisfied if it is satisfied in the early universe since $s\,>\,2\,>\,\frac{1}{3}$.)
Hence Eq.(\ref{t6}) becomes
\begin{equation}
\frac{\dot{C}_\chi}{a^3}\,=\,-\beta\,C_{\chi}^2\,\sigma_0\,v_0\frac{a^{(5s+1)}}{a^6}
\label{t12}
\end{equation}
which, in turn, implies
\begin{eqnarray}
\frac{dC}{C^2}&=&-\beta\,\sigma_0\,v_0\frac{a^{(5s+1)}}{a^3}dt\,=\,-\beta\,\sigma_0\,v_0\,a^{5s-2}\frac{da}{\xi\,a^{(1-s)}}
\nonumber \\
&&\,=\,-\frac{1}{\xi}\beta\,\sigma_0\,v_0\,a^{6s-3}\,da \;.
\label{t13}
\end{eqnarray}
After integrating out (\ref{t13}) we have
\begin{equation}
C_\chi\,=\,
\frac{C_1}{\frac{C_1\beta\sigma_0v_0}{(6s-2)\xi}\left(\,a^{6s-2}-a_1^{6s-2}\right)+1} \;.
\label{t15}
\end{equation}
Hence we obtain
\begin{equation}
C_\phi\,=\,C_1\,-\,
\frac{C_1}{\frac{C_1\beta\sigma_0v_0}{(6s-2)\xi}\left(\,a^{6s-2}-a_1^{6s-2}\right)+1} \;.
\label{t16}
\end{equation}
An analysis similar to the case i) in Section III. B. for initial times in this case leads to
\begin{equation}
C_\phi\,\simeq\,
\frac{C_1\beta\sigma_0v_0}{(|6s-2|)\xi}\,a^{|6s-2|}\left[\,1-\left(\frac{a_1}{a}\right)^{|6s-2|}\right] \;,
\label{t16a}
\end{equation}
where the condition $s\,>\,2\,>\,\frac{1}{3}$ is imposed to exclude the tachyonic case. At late times (\ref{t16}) may be approximated by
\begin{equation}
C_\phi\,\sim\,C_1\,-\,
\frac{(|6s-2|)\xi}{\beta\sigma_0v_0}\,a^{-|6s-2|}~~~\mbox{for}~~6s-2\,>\,0~\mbox{if}~\frac{C_1\beta\sigma_0v_0}{(|6s-2|)\xi}\,a^{|6s-2|}\,\gg\,1 \;.
\label{t16c}
\label{t16d}
\end{equation}
(\ref{t16c}) implies that if the processes $\chi\chi\,\rightarrow\,\phi\phi$ continue till very late times then $C_\phi$ may reach its maximum value $C_1$. However the conclusions for late times here are not wholly reliable since we have neglected the effect of statistics and the effect of the processes $\phi\phi\,\rightarrow\,\chi\chi$ which can not be neglected while they can be neglected at initial times. Eq. (\ref{t16c}) is reliable only if the number density of $\phi$s has not reached a large value at late times yet.

\end{document}